# ROBUST TIMING SYNCHRONIZATION FOR AC-OFDM BASED OPTICAL WIRELESS COMMUNICATIONS

*Bilal A. Ranjha, Mohammadreza A. Kashani, Mohsen Kavehrad, and Peng Deng, The Pennsylvania State University, University Park, PA*

## Abstract

Visible light communications (VLC) have recently attracted a growing interest and can be a potential solution to realize indoor wireless communication with high bandwidth capacity for RF-restricted environments such as airplanes and hospitals. Optical based orthogonal frequency division multiplexing (OFDM) systems have been proposed in the literature to combat multipath distortion and intersymbol interference (ISI) caused by multipath signal propagation. In this paper, we present a robust timing synchronization scheme suitable for asymmetrically clipped (AC) OFDM based optical intensity modulated direct detection (IM/DD) wireless systems. Our proposed method works perfectly for ACO-OFDM, Pulse amplitude modulated discrete multitone (PAM-DMT) and discrete Hartley transform (DHT) based optical OFDM systems. In contrast to existing OFDM timing synchronization methods which are either not suitable for AC OFDM techniques due to unipolar nature of output signal or perform poorly, our proposed method is suitable for AC OFDM schemes and outperforms all other available techniques. Both numerical and experimental results confirm the accuracy of the proposed method. Our technique is also computationally efficient as it requires very few computations as compared to conventional methods in order to achieve good accuracy.

## Introduction

In recent years, optical wireless (OW) communications especially intensity modulation direct detection (IM/DD) systems have been receiving an increasing attention from researchers around the globe due to availability of high speed optical transmitters such as light emitting diodes (LEDs). Employing these transmitters, optical signal can be used as a carrier for information transmission. Optical intensity can be easily modulated with baseband electrical signals and can be detected using optical detectors at the receiver. Optical transmitters such as LEDs operating in both visible and non-visible spectrum can be used for IM/DD communications [1-11].

RF-restricted environments such as aircraft cabins are considered as one of the candidate areas for OW communication [12]. However, due to multipath nature of indoor environments, efficient communications schemes need to be addressed which can combat multipath distortion and intersymbol interference (ISI) [13] caused by multipath signal propagation. OFDM has been used in many radio frequency (RF) communication standards such as LTE, DVB-T and WiFi, due to its ability to mitigate signal distortions caused by multipath reflections. However, due to high sensitivity of OFDM to carrier frequency offset and timing synchronization errors, efficient timing synchronization and carrier frequency offset correction techniques need to be used at the receiver for RF based systems. For IM/DD based OFDM systems, timing synchronization errors can cause performance degradation and therefore requires an efficient timing synchronization scheme that has high accuracy and is computationally efficient. It should also be a generic scheme not tailored for a specific technique.

A large number of papers have been published on timing synchronization schemes for RF based OFDM systems [14-18]. These techniques are not directly applicable to OFDM based intensity modulated direct detection (IM/DD) systems because of the unipolar nature of output signal. Therefore, timing synchronization schemes that are suitable for IM/DD systems need to be addressed. In this paper, we focus on timing synchronization for asymmetrically clipped (AC) based OFDM systems. Recently, in [19] Tian *et al.* proposed a technique tailored specifically to ACO-OFDM. This scheme may not work for other AC systems and the detection accuracy of this scheme also depends on the choice of training symbol used. Some training symbols may not give perfect accuracy even at high SNR without noise and multipath. In [20], authors present a method that utilizes symmetry of ACO-OFDM time domain output symbol with some additional redundancy. However, the channel cannot be estimated using this technique.

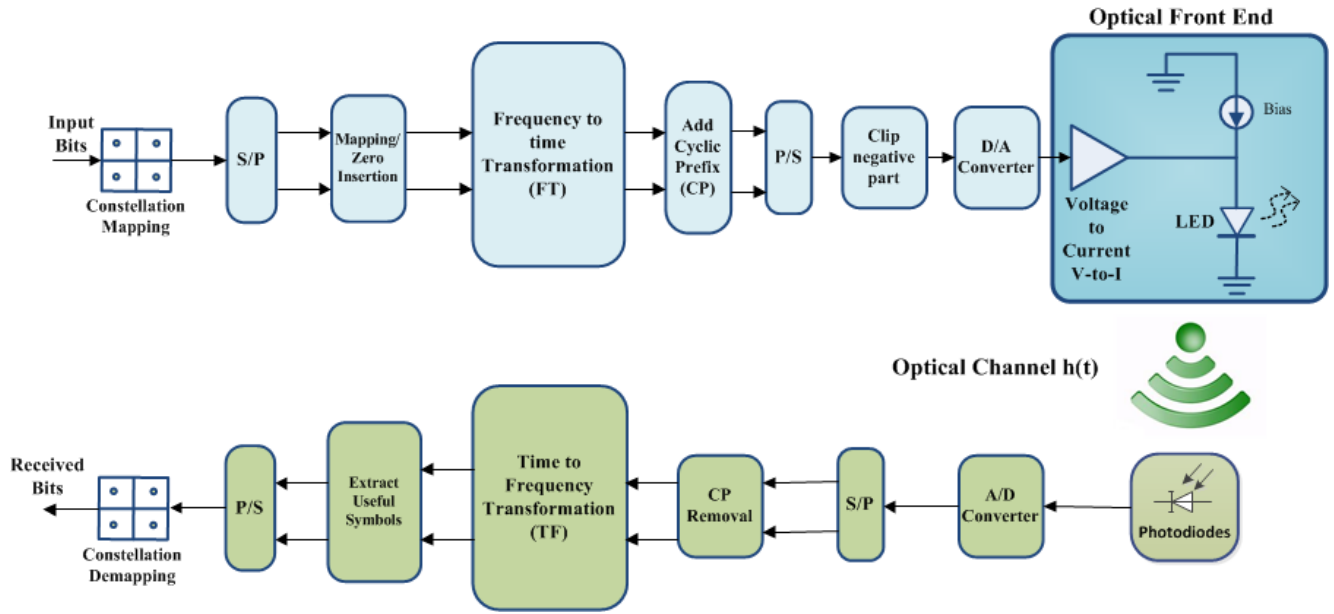

**Fig. 1. A generalized block diagram of asymmetric clipped based OFDM systems**

In this paper, we present a novel and robust timing synchronization method that works perfectly for all AC systems, namely asymmetrically clipped optical OFDM (ACO-OFDM), PAM-modulated discrete multitone (PAM-DMT) and discrete Hartley transform (DHT) based optical OFDM, and can also be used for channel estimation simultaneously. Our technique not only gives the best performance, but due to flexibility in size of correlation length, we can achieve perfect accuracy even with smaller correlation length and at lower SNR.

## OFDM based OW systems

In RF based OFDM systems, output signal is bipolar and complex. This signal cannot be easily transmitted in an OW system since light intensity cannot be negative, and we cannot transmit a complex signal using a single optical transmitter such as LED [21]. Therefore, output OFDM signal has to be made real and positive to make it suitable for optical transmission. Hermitian symmetric input data to OFDM block generates a real output signal. However, to make signal positive several OFDM schemes have been proposed for (IM/DD) OW systems. Among them, one is called DC-Biased OFDM [8] wherein a DC bias is used to make the output signal positive. Other schemes involve clipping negative part of the output signal. PAM-DMT [22] is one of these clipping based schemes where the complex part of each subcarrier is modulated with a real symbol resulting in clipping noise to fall on the real part of the same subcarrier. Another clipping based scheme known as asymmetrically clipped optical OFDM (ACO-OFDM) uses only odd subcarriers modulated by complex constellation symbols [23-24]. This results in clipping noise to fall only on even subcarriers. Therefore, in both clipping based strategies, the clipping noise is always orthogonal to the transmitted symbols enabling easy recovery of the desired data at the receiver. Another technique called discrete Hartley transform (DHT) based optical OFDM [25] uses real input symbols and generates a real bipolar output signal using Hartley transform. The characteristics of output signal are similar to those in ACO-OFDM.

In this paper, we will only focus on these three AC based OFDM techniques. A generic block diagram of AC based OFDM system is shown in Fig. 1. Only constellation mapping, mapping and zero insertion, frequency to time transformation (FT), time to frequency (TF) domain transformation and extract symbols block will perform different operations on the input data for each scheme. Rest of the transmitter and receiver blocks will remain same.

## ACO-OFDM

In this scheme, input vector consists of $M = N/4$ complex symbols drawn from a complex 2D constellation mapping scheme such as m-ary quadrature amplitude modulation (M-QAM) which will modulate only odd subcarriers in the first half of $N$ subcarriers. $N$ is the total number of subcarriers available and is equal to the size of IFFT. In ACO-OFDM, FT will perform IFFT operation on input data. The conjugate of these symbols modulates the odd subcarriers of second half of $N$ subcarriers to meet the Hermitian symmetry requirements. Therefore, the input data vector is transformed to

$$\mathbf{X} = \left[0, X_0, 0, X_1, 0, \ldots, X_{N/2-1}, X^*_{N/2-1}, 0, \ldots, X^*_0\right] \quad (1)$$

where $X_k$ is the complex symbol. The first (DC) and $N/2^{nd}$ subcarriers are set to zero to obtain a real output signal. The time domain output signal is generated by taking the IFFT of the input vector

$$x_n = \frac{1}{N} \sum_{k=0}^{N-1} X_k \exp\left(j 2\pi \frac{k}{N} n\right) \quad (2)$$

A cyclic prefix (CP) is added to this discrete time output signal. This bipolar signal is asymmetrically clipped by clipping negative part and passed through digital-to-analog (D/A) converter to generate a continuous time domain signal and ultimately modulates the intensity of the optical transmitter such as LED. Clipping noise generated by clipping negative half of time domain signal falls only on even subcarriers. Therefore, the transmitted symbols are not affected by clipping noise which enables easy recovery of transmitted data at the receiver.

At the receiver, an optical detector converts the intensity into an electrical signal. This signal is corrupted by electronic noise generated by the electronic components and the ambient noise from the surrounding light sources. This noise is usually modeled as additive white Gaussian noise (AWGN). The noise corrupted signal is then passed through an A/D converter to generate a discrete time signal. After removing CP, the TF block performs $N$-point FFT operation on the input discrete time samples. The noise corrupted constellation symbols are extracted from FFT output and de-mapped to generate the output bits.

## PAM-DMT

In this OFDM based scheme, $N/2$ symbols drawn from a real mapping scheme such as m-ary pulse amplitude modulated (M-PAM) are used to modulate the complex part of each subcarrier. However, the DC and $N/2^{nd}$ subcarriers are not modulated to fulfill the Hermitian symmetry requirements. Therefore, the data vector forming the input to FT block will be

$$Y = \left[0, Y_0, Y_1, Y_2, \ldots, Y_{N/2-1}, 0, Y^*_{N/2-1}, \ldots, Y^*_1, Y^*_0\right] \quad (3)$$

where $Y_k = ib_k$ and $b_k$ is the real valued symbol drawn from the real constellation, e.g. M-PAM. In PAM-DMT, FT will perform IFFT operation on input data. The real part of each subcarrier is not modulated. Remaining front end blocks perform the same operation on this bipolar signal as that in ACO-OFDM.

According to [22], by only modulating complex part of each subcarrier, clipping noise falls on real part of each subcarrier and leaves imaginary part undisturbed. Therefore, clipping operation does not affect transmitted symbols and hence can be easily recovered at the receiver.

The receiver architecture of PAM-DMT is similar to that described in ACO-OFDM. However, after FFT block, only imaginary part of each output symbol is extracted to recover the estimated PAM symbols.

## DHT-OFDM

In DHT based optical OFDM, a vector of length $N/2$ of real symbols drawn from a real constellation such as M-PAM forms input to the FT block. In this scheme, FT block will perform inverse fast Hartley transform (IFHT). According to [25], if the input symbols only modulate odd indexed subcarriers, clipping noise will only fall on even indexed subcarriers. Therefore, the input vector of length $N$ is transformed to

$$\mathbf{X} = \left[ 0, X_0, 0, X_1, 0, \ldots, X_{N/2-1}, X^*_{N/2-1}, 0, \ldots, X^*_0 \right] \quad (4)$$

by zero insertion block. However, we do not need conjugate of the input symbols since IFHT is a real transform and will generate real signal with real input symbols. Therefore, the length of useful input symbols is $N/2$. An N-point IFHT is performed on $\mathbf{X}$ to output a real bipolar signal

$$x(n) = \frac{1}{\sqrt{N}} \times \sum_{k=0}^{N-1} X(k) \left[ \cos\left(2\pi \frac{k}{N} n\right) + \sin\left(2\pi \frac{k}{N} n\right) \right] \quad (5)$$

Remaining front end blocks perform the same operation on this bipolar signal as that in ACO-OFDM and finally transmit it using an optical transmitter.

At the receiver, reverse operation is performed to recover transmitted bits. After removal of CP, fast Hartley transform (FHT) is performed by TF block on the received signal which outputs estimated transmitted symbols. DHT has a self-inverse property which enables us to use the same software routines as used by transmitter.

## Conventional timing synchronization techniques

In this section, we briefly discuss three previously proposed timing synchronization methods. Two of these techniques were proposed for RF based OFDM and one for ACO-OFDM.

### Schmidl's method

In [16], Schmidl presented a timing synchronization method based on autocorrelation of two identical halves of OFDM training symbol. Such a training symbol can be generated by modulating only even subcarriers with complex constellation symbols such as M-QAM. The resulting time domain training symbol of length $N$ will have two repeated halves $[\mathbf{A}, \mathbf{A}]$ excluding CP where $\mathbf{A}$ represents the first $N/2$ samples of output symbol. The timing metric used in Schmidl's method suffers from a plateau due to CP which results in some uncertainty in start of the training symbol. Therefore, this technique will not predict start location of frame very accurately.

### Park's method

In [17], Park's proposed a timing synchronization scheme ensuring a sharp peak in the timing metric to precisely indicate start of training symbol. To achieve this, a new time domain training symbol was used which can be generated by modulating only even subcarriers with real valued random symbols such as M-PAM. Resulting time domain symbol will have a format $[\mathbf{A}, \mathbf{B}, \mathbf{A}^*, \mathbf{B}^*]$ where $\mathbf{A}$ represents first $N/4$ samples of this time domain training symbol and $\mathbf{B}$ represents mirror image of $\mathbf{A}$. This method results in several sharp peaks one of which will occur at the correct location of start of training symbol.

The above mentioned schemes cannot be directly applied to AC optical OFDM systems. Therefore, a modified version [15], wherein complex constellation symbols satisfying Hermitian symmetry are used as input to IFFT block. In the modified version, Schmidl's timing metrics shows a flat region during the length of CP of training symbol. However, Parks method does not have this flat region but has four distinct peaks one of which is at the correct timing instant. We will use this modified version of Park's method in our paper for comparison and analysis.

A plot showing average of the timing metric for Schmidl's and parks method with modified training symbols is shown in Fig. 2. We can see that Schmidl's timing metrics shows a flat region during the length of CP of training symbol. However, Parks method does not have this flat region but has four distinct peaks one of which is at the correct timing instant.

### Tian's method

Recently Tian [19] proposed a timing synchronization method tailored to ACO-OFDM system. In this technique, a new time domain training symbol is used that has a format,

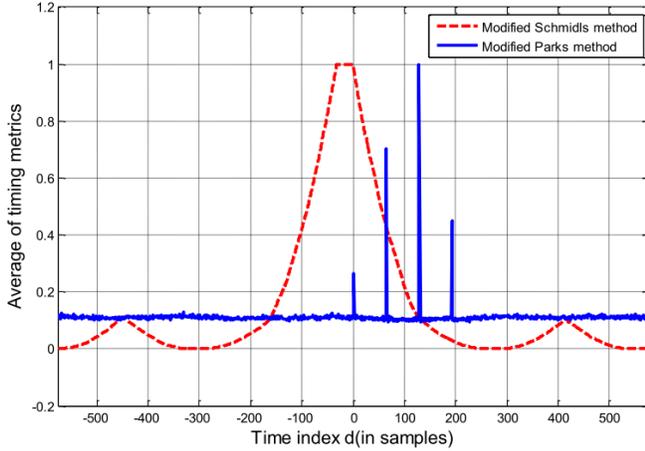

Fig. 2. Average of Schmidl's and Park's timing metrics with modified training symbol suitable for ACO-OFDM in the absence of AWGN and multipath

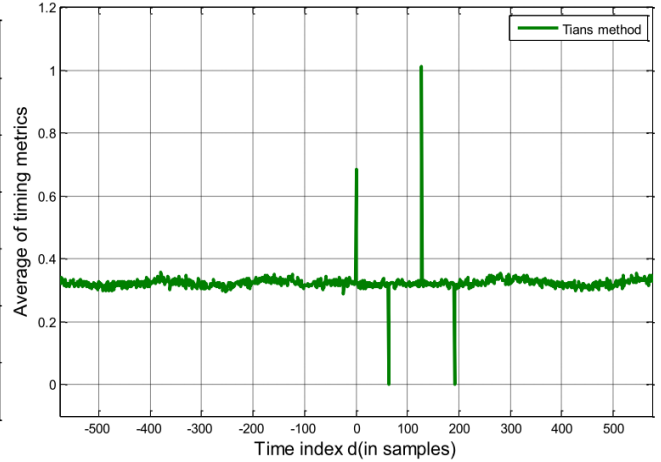

Fig. 3. Average of Tian's timing metrics in the absence of AWGN and multipath

.

$$\mathbf{X} = [0\ \mathbf{C}\ 0\ (-\mathbf{C}^{mirror})_{clip}\ 0\ (-\mathbf{C})_{clip}\ 0\ (\mathbf{C}^{mirror})_{clip}]  \quad (6)$$

where $\mathbf{C}$ represents $N/4-1$ samples of output training symbol. Such a training symbol can be produced by modulating odd subcarriers with real constellation symbols and even subcarriers by zero. The author presented several timing metrics but we will only present one metric for analysis and comparison. This metric known as simple timing metric is given by

$$M(d) = \frac{1}{N/8-1} \sum_{n=1}^{N/4-1} r(d-n)r(d+n) \quad (7)$$

Since we are using total average electrical power of unity for our analysis and comparison of various timing synchronization schemes, therefore we will be using a factor of $N/8-1$ in the numerator of the above metric. Fig. 2 shows average of simple timing metric. To generate results in Fig. 3, a total of 10,000 random training symbols were used with IFFT size of $N = 256$ and CP length of $N/8$. To get more realistic results, each training symbol was followed and preceded by another random ACO-OFDM symbol. The figure shows that besides the main peak at the correct timing instance of $d = N/2$, there is another peak at $d = 0$. The difference between these two peaks is not high which will reduce correct detection probability especially at low SNR.

## New Timing synchronization for AC based OFDM system

In this section, we present a new timing synchronization scheme that can be used for all AC OFDM systems. Although the fundamental approach used is the same for all systems, some minor modifications are required to make it suitable for each. The details are presented below.

### Symbol timing estimation for ACO-OFDM

Our proposed method uses very important property of ACO-OFDM output waveform which has a format $[\mathbf{C}_{clip}\ \mathbf{D}_{clip}]$ where $\mathbf{C}$ represents the first $N/2$ samples and $\mathbf{D} = -\mathbf{C}$ represents negative part of first $N/2$ samples of unclipped output time domain ACO-OFDM symbol. $\mathbf{C}_{clip}$ and $\mathbf{D}_{clip}$ represent the first and the second half of the clipped output symbol. This shows that the negative parts of the first $N/2$ samples of unclipped time domain symbol are present in the second half of $N/2$ samples of clipped symbol. Therefore, we can easily reconstruct a bipolar signal of length $N/2$ with these two halves that will be identical to the original unclipped bipolar signal of length $N/2$. The bipolar signal is constructed as

$$\mathbf{r}_{BP}(n) = \mathbf{C}_{clip}(n) - \mathbf{D}_{clip}(n) \quad (8)$$

where $0 \leq n \leq N/2 - 1$ and subscript BP represents bipolar. This reconstructed bipolar signal can be used to perform correlation with a local copy of training symbol $p(n)$ known at the receiver to correctly estimate starting location of OFDM symbols. We will use timing metric given by

$$M(d) = \frac{1}{L} \sum_{n=0}^{L-1} r_{BP}(n+d) p(n+d), \quad L = 1, 2, \ldots, N/2 \quad (9)$$

where $r_{BP}(n)$ is the reconstructed bipolar signal. $L$ is the cross-correlation length that can be set based on the desired performance. A higher value of $L$ offers a better performance. Maximum of this timing metric will be used to find the starting location of OFDM training symbol. We assume throughout this paper that average output electrical power of ACO-OFDM output training symbols before clipping is unity i.e. $E\{p^2(n)\} = 1$.

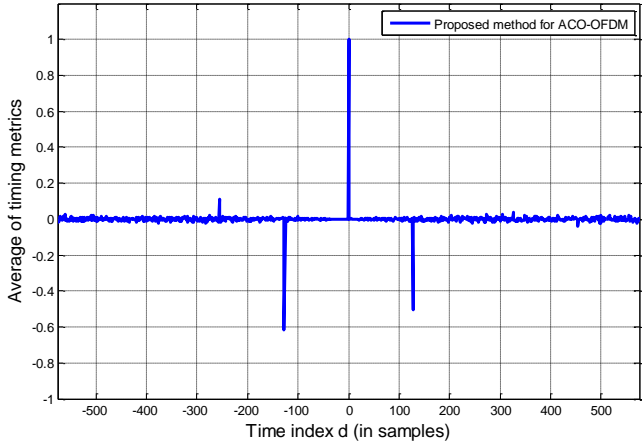

**Fig. 4. Average of proposed timing metrics in the absence of AWGN and multipath for ACO-OFDM.**

A plot of average of this timing metric with $L = N/2$ for ACO-OFDM is shown in Fig. 4. To generate these results, a random sample of 10,000 training symbols was used preceded and followed by a random ACO-OFDM symbol with CP. From the figure, we can clearly see that a peak occurs at the correct location at $d = 0$. There are two other negative peaks at $d = -N/2$ and $d = N/2$ occurring before and after the main peak, respectively. These are caused by the negative correlation of first and second half of local signal with the received signal. Since we are using maximum of the timing metric, we will ignore these peaks as they will not cause any uncertainty in the correct location identification. There is also a small peak occurring at $d = -N$. This is due to the correlation of local training symbol with CP of received training symbol. The magnitude of this peak depends on the size of CP. Since CP length is usually small compared to the length of useful part of symbol, the magnitude of this peak will be small compared to the main peak and thus will not cause any uncertainty in correct location of beginning of training symbol and will not result in erroneous detections.

*Symbol timing estimation for PAM-DMT*

In PAM-DMT, the output waveform has the format $[0 \; \mathbf{C}_{clip} \; 0 \; \mathbf{D}_{clip}^{mirror}]$ where $\mathbf{C}$ represents the first $N/2 - 1$ samples excluding the first sample and $\mathbf{D} = -\mathbf{C}$ represents negative of first $N/2 - 1$ samples of unclipped PAM-DMT output symbol. $\mathbf{C}_{clip}$ and $\mathbf{D}_{clip}$ represent clipped version of $\mathbf{C}$ and $\mathbf{D}$, respectively. In this case, the second half of output symbol contains a mirror image of negative samples of the first half. Therefore, the bipolar received signal can be reconstructed by

$$\mathbf{r}_{BP}(n) = \mathbf{C}_{clip}(n) - \left(\mathbf{D}_{clip}^{mirror}(n)\right)^{mirror} \quad (10)$$

where $0 \leq n \leq N/2 - 1$. This bipolar signal will be correlated with a local copy of training symbol to locate the beginning of OFDM training symbol. We will use maximum of the same timing metric used by ACO-OFDM given in (9) with the bipolar signal reconstructed using (10).

A plot of average of this timing metric for correlation length $L = N/2 - 1$ is shown in Fig. 5.

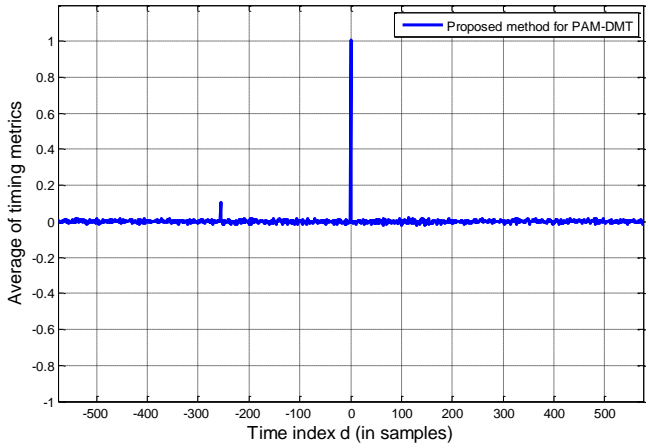

**Fig. 5.** Average of proposed timing metrics in the absence of AWGN and multipath for PAM-DMT

From the figure, we see that there is one peak at $d = 0$ that shows correct location of start of OFDM training symbol. There is another small peak at $d = -N$. This is due to the correlation of local training symbol with the CP of received training symbol. The properties of this small peak are similar to those described in case of ACO-OFDM. Due to high difference in the magnitude of these two positive peaks, we expect a high probability of correct detection compared to other previously proposed techniques.

*Symbol timing estimation for DHT based O-OFDM*

In DHT based O-OFDM, output waveform has the same format as that of ACO-OFDM, i.e. $[\mathbf{C}_{clip}\ \mathbf{D}_{clip}]$. Therefore, at the receiver, we will reconstruct a bipolar signal in the same way as was reconstructed in ACO-OFDM. Same timing metric will be used for DHT based O-OFDM system given in (9). A plot of average of this timing metric for $L = N/2$ is identical to that obtained for ACO-OFDM. Therefore, to avoid repetition of results, we will not show results for DHT based O-OFDM systems as they will be identical to ACO-OFDM.

## Experimental results

An experimental test-bed is the perfect setup to verify the results discussed earlier. We employ USRP210, a software defined radio (SDR), as the primary hardware and software interface for the test-bed. The experimental setup is shown in Fig. 6. As can be seen, the baseband data from the host PC properly processed goes through the first USRP210 kit, where it is converted to RF signal. Instead of feeding this signal to an antenna, it is fed to a driver circuit, which in turn drives the LED transmitter which is an OSRAM OSTAR Phosphor White LED. The LED transmitter is thus properly modulated by the data from the host PC and the light propagates through the channel. The intensity of the light is detected by the Si PIN photodiode, which produces a current signal proportional to this intensity. The signal is then amplified by an amplifier and fed to a second USRP210 that sends it to the target PC as baseband data. The data is further processed at the target PC.

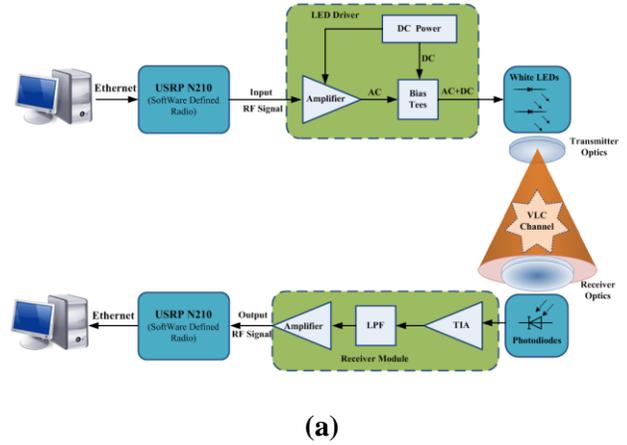

(a)

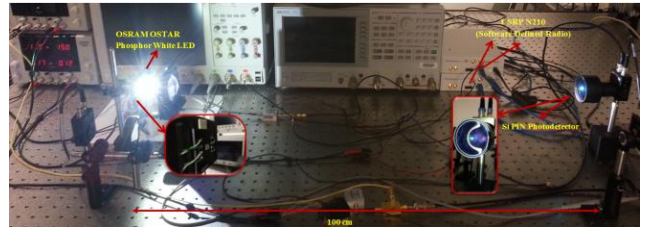

(b)

**Fig. 6.** a) Schematic of the experimental setup. b) Real implementation with software defined radio systems.

Fig 7.a-d demonstrates the timing metric for the bipolar correlation method with $L = N/2$ and CP length of $N/8$ where $N = 256, 512$. The symbols are drawn from 4-and 16-QAM constellations. As can be clearly observed, the experimental results validate the simulation results discussed earlier.

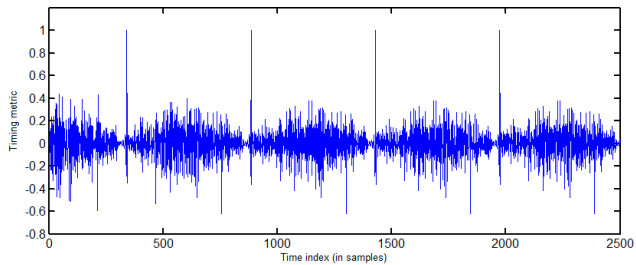

**(a)**

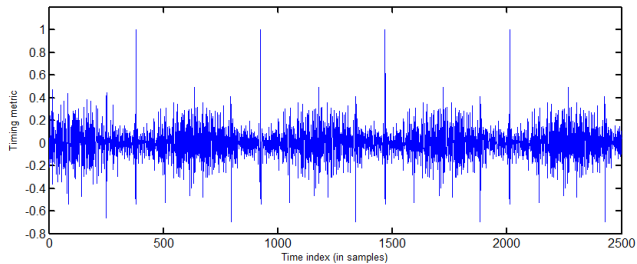

**(b)**

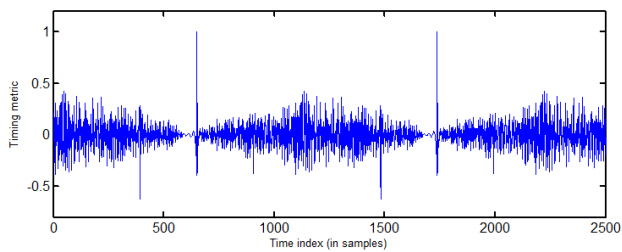

**(c)**

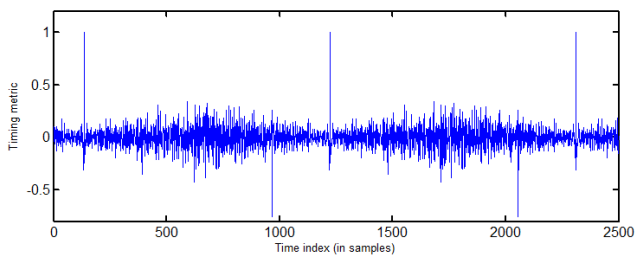

**(d)**

**Fig. 7** Average of timing metrics for bipolar correlation method for consecutive ACO-OFDM symbols with a) $N = 256$ and 4-QAM modulation b) $N = 256$ and 16-QAM modulation c) $N = 512$ and 4-QAM modulation b) $N = 512$ and 16-QAM modulation.

## Conclusions

In this paper, we have presented a novel timing synchronization scheme that can be used to estimate symbol timing for asymmetric clipping based OFDM systems using IM/DD. This timing synchronization scheme can be applied to ACO-OFDM, PAM-DMT and DHT based optical OFDM systems. Unlike other timing synchronization schemes, no special format of the training symbols is used. Instead, regular OFDM symbols generated by asymmetric clipping schemes are used. Our timing metric uses correlation of a local copy of the training symbol with a bipolar signal reconstructed from unipolar received signal. The bipolar signal can be easily constructed from unipolar signal due to the fact that output signal of asymmetric clipping schemes carry positive and negative parts of first $N/2$ samples of time domain signal. Simulations and experimental results have been presented to confirm the accuracy of the proposed method.

## Acknowledgements

The authors would like to thank the National Science Foundation (NSF) ECCS directorate for their support of this work under Award #1201636, as well as Award #1160924, on the NSF "Center on Optical Wireless Applications (COWA–http://cowa.psu.edu"

## Email Address

bar5254@psu.edu

mza159@psu.edu

mkavehrad@psu.edu

pxd18@psu.edu